# Non-Parametric Bayesian Rejuvenation of Smart-City Participation through Context-aware Internet-of-Things (IoT) Management

Rossi Kamal, Choong Seon Hong
Department of Computer Engineering,
Kyung Hee University, South Korea
E-mail:rossikamal@rossikamal.info,cshong@khu.ac.kr

## Abstract

Tweaking citizen participation is vital in promoting Smart City services. However, conventional practices deficit sufficient realization of personal traits despite socio-economic promise. The recent trend of IoT-enabled smart-objects/things and personalized services pave the way for context-aware services. Eventually, the aim of this paper is to develop a context-aware model in predicting participation of smart city service. Hence, major requirements are identified for citizen participation, namely (a) unwrapping of contexts, which are relevant, (b) scaling up (over time) of participation. However, paramount challenges are imposed on this stipulation, such as, un-observability, independence and composite relationship of contexts. Therefore, a Non-parametric Bayesian model is proposed to address scalability of contexts and its relationship with participation. Finally, developed systematic prototype pinpoints major goals of context-aware application from participants' opinions, usage and feedback.

## Introduction

World has speculated a huge penetration on urbanization over last decade, which is anticipated to be continued, even with a higher rate. This global trend of urbanization is noticed from the statistical exposition, 'Half of world population is living in cities in 2013, whereas the ratio will be reached by in Asia and Africa around 2020 and 2035, respectively. Last but not the least, global urban population will be nearly doubled (i.e. 3.6 to 6.3 million) by 2050'[1] . Consequently, there is an ever increasing demand of sustainable city, with improved city-governance and quality of urban-life[2].

Cities are reminiscent of verdict, culture and finance from primeval ages. They have been medial plane for cultural or commercial-interchange and public-governance. However, contemporary urbanization-trend are striving us to apply technology for the amelioration of socio-economic conjecture. Hence, recent Information Communication Technologies (ICT) with the advent of information and operational management are going to play a pivotal role in urbanization.[1] [2]

Smart City is aimed at undertaking networked information for operational management of an urban-life. It comprises of services that use, among others, Smart-devices, wearable sensors, intelligent vehicles around the city, enabling accumulation of monitoring information, to react autonomously in real time, with less or without human intervention. Operational management comprises of configuration management, green computing, load balancing, quality-of-protection, disaster-response, to name a few. However, information is regarded as vital factor in multidimensional challenges, such as sharing, transfer and analysis of knowledge-base.[3][4][5]

An abrupt paradigm shift is noticed in urban space among internet-business community with the replacement of infrastructure providers by operators and service providers. Accordingly, groundbreaking promise of crowd-intelligence pings a transition from straightforward data pipe towards optimized Big data. ec not the least, contextual aspects, stemmed from increasing demand of personalization, necessitate leveraging advanced operations, such as, fusion, cleansing and quality-of-protection.

Recently the proliferation of IoT-enabled devices[6] and large-scale adoption of personalized services[7] are striving innovative solutions towards societal challenges with everenhanced capacity of Smart-devices. Having strong potential impact on both formal and informal spaces, such as quality of experience and urban-provisioning, technology is being considered as key player to bring Smartness in community. However, intelligence, sensitivity and responsiveness embedded in things or objects, or ubiquitous surrounding, are demanding redefinition of urban space, ecosystem, and even measurement approaches.

Enabling smart-city information as a utility back social ecosystem. It includes exploration of value in massive data, which is collected from heterogeneous sources, such as agencies(i.e. public or private) or social interaction (i.e. Facebook or Twitter) or regular-usage (i.e. YouTube). Since most data is location based, geography is regarded as common platform for inferring value from multiple sources. However, sometimes, not only geography, but also other essences of data, such as mobility, consumer-behavior, usage-hour(peak/off-peak), are regarded important. Uncertainties and privacy concerns over data from heterogeneous sources necessitate high control over overall monitoring information. In this context, analytic is indispensable for unwrapping underlying structure for optimal decision making citeyinhai. Thereafter, 'smart city' is often coined by 'automated city', resembled by on-demand provision of participation for both regulatory and emergency services in completely automotive manner[3][4]

The objective of the work is to develop instinctive methodology, where citizen participation is promoted with adequate quantification of contexts. Hence, key prerequisites are spotted, which automation should clinch. Primarily, a wrapping model is essential, which accurately annotates latent theme from plethora of participatory information. However, the latter stage demands scalability, which illuminates transformation over time and attachment on connected themes. This provides both ease and convenience in optimized decision making for automated service-provisioning.

To address above challenges, our solution predicts an NonParametric Bayesian[8] autonomous model for unwrapping themes and scaling up in urban living. Medium-scale assessments (i.e. subjective, objective and acceptability-driven) are conducted to quantify statistical methodology in inferring polychromatic experiences of participants. However, an impractical implication of statistical information often leads to conclusion, that is not apotheosis. Hence, our systematic prototype attempts to grapple with some completely baffling ingredients.

A synopsis of our work is as follows, (a) A methodical highlight of critical complications of smart city, (b) Nonparametric Bayesian confrontation of unwrapping of contexts and then participation of citizens, (c) Assessment on implication of developed machine-learning scheme on medium-scale environment, (d) Concluding remarks

**Problem Formulation**
Motivated by global alert on urban revamp, technologists and policy decision makers are taking sustainable plans individually. Technologists have stepped in with pragmatic approaches (e.g. Microsoft' CityNext platform, Siemens's Sustainable city) envisioning 'smart planet and smart cities', through adequate usage of global ICT. Thus, smart city market is predicted to be nearly hundred billion of dollars by 2020, with annual rate of nearly 16 billion. Joint initiatives are taken by ICT industries (e.g. IBM, CISCO) and public/public organizations to improve urban life in different areas, such as transportation, energy, governance, etc. Large-scale smart city projects are launched by alternative secondary markets, such as, Europian Union and relevant local initiatives. Europian digital agenda comprises of projects, such as European Digital Cities,1 InfoCities,2 Intel City roadmap, and EUROCITIES. FuturICT, is EU's long-term initiative, which sustains resilience in Smart city through adequate reasoning and management of socio-global complex infrastructure.

However, different techniques are adopted by different cities, who are involved in the venture. Four major areas, such as living, working, mobility, public-space, are emphasized by Amsterdam authority to converge into a real smart city in near future by reducing $CO_2$ emission and renovating public, transportation and urban spaces. London city authority is experimenting to enhance urban life by managing traffic flows, water supplies and extreme weather conditions. South Korea is developing Songdo Business District, a smart city without carbon-emission. In this context, the geographic dispersion and entailed complex technologies are reckoned as major barriers for decision makers, who are sensing a scarcity of mass participation, which demonstrate the feasibility of real-scale deployment. Investors are struggling to reach a risk-free solution, which help them to aggregate small and medium-scale project into a large-scale participation. The adversities are expected to be more complex with required interaction of private and public initiatives.

Hence, our overall objective is to devise systematic methodology, such that participation is stimulated by accurate quantification of citizen's usual contexts, for example, emotion (i.e consumer-behavior), location (e.g. hospital, public-space, etc.), social-presence (e.g. social network, etc.).

However, unwrapping common contexts from plethora of geographically dispersed and technologically-complex knowledge-base (e.g. Unwrapping problem) is a difficult task. Conventional schemes involve manual configuration by service providers, which is both time and money consuming.

Moreover, representation of knowledge-base (e.g. Involvement Problem) based on latent contexts, is also cumbersome. The instrumentation requires quantification, whether information is matched to that contexts, which scale up over time and are tied within each others

## Methodology

A hypothetical conjecture of context-aware skeleton is visualized by dissecting urban participation information. Annotated contexts should be unveiled from citizen database at inaugural phase. Complex association among contexts, which evolve over time, should be dealt adequately at latter stages. Therefore, uncovering contexts and then scaling up over time speculate a Non-parametric Bayesian model, which is amenable to statistical inference. Participation knowledge-base is deemed to be ameliorated amidst a generative process, which involve citizen contexts. This generative process is conventionally represented by joint probabilistic distribution over observed information and latent random variables. Hence, the joint distribution is defined in terms of conditional distribution of latent contexts, given citizen information. Eventually, the conditional distribution of the latent variables, given the knowledge-base turns to be a posterior distribution. Thus, the generative process involves joint distribution of citizen context and participation information. In general, the posterior distribution resembles the original problem for finding latent variable from observed information.

Therefore, in our Smart-city use-case, this is expressed by conditional distribution of contexts, given citizen knowledge -base. Ultimately, the posterior is sum of joint distribution over all possible instances of latent variables. Owing to this statistical facet, the numerator represents joint distribution of all random variables for any latent variable. Moreover, the denominator comprises of marginal probability distribution of latent variables, which resemble probability of corpora in devised scheme. Finally, rejuvenation of Smart City participation is intractable and thereby amenable to statistical inference.

Because, the above posterior is not computable due to the denominator, which is exponential large. Eventually, context prediction is turned to intractable. Hence, citizen's participation estimation is amenable to statistical inference, which approximates suitable probabilistic model using citizen knowledge base.

## Assessments

Our subjective assessment contemplates 71 participants perceptions for different contexts (i.e. emotion, weather, location) in Smart-device-usage, Radio/TV and social-network services.Then, smart-device usage-history (e.g. device id, application-genre, service-name, time) is collected from lab members on different parts of day. Moreover, convenient Smartphone Graphical User Interface (GUI) is developed to accept participants real-time feedback during and after application-usage. Finally, in group validation, participants are requested to rate quality of stimuli with dichotomous choice. Due to space limitation, readers are encouraged to refer to our developed systematic platform[7].

### A. Subjective Assessments

Opinions are requested in a way, such that participants never turn confused (Table I, II). Hence, questionnaires are kept as simple as possible, which does not seem abstract to participants. In this context, several brainstorming sessions are undergone among co-authors (i.e. project members) to sustain a rhythm in the survey lifespan. The overall objective is to engage complete concentrations of participant, who promises to accomplish it with full motivation

### B. Objective Assessments

Objective assessment(Table III) entrenches high level of control in laboratory environment. Its archetypal outcome is rendered through communication patterns of individuals or groups. Participants use-profile validates this outcome to construct suitable models and parameters. These models are worthwhile to standardize methodologies for comparative analysis in real-life experiment.

### C. Acceptability-based Assessments

Acceptability (Table IV) measures participants overall experience on developed applications in online fashion. Subjective and objective assessments are conducted in limited space (i.e. subjects from only four communities) and indoor environment (I.e laboratory set-up), respectively. As a result, participants perception and satisfaction are seldom reflected, especially in real usage contexts. Hence, acceptability accumulates real-time feedback of participants in practical application scenarios. Thus, network, device and content-centric features are experienced in different dimensions through online and collaborative feed-backs

### D. Validation of Conducted Assessments

Cognitive gap between So So, Worst is different than that between Awesome and Good. Acceptability (Table V) is rated according to arithmetic average of ratings, which is not applicable in ordinal scale measurements. Such precision limitations are overcome with dichotomous choice based easier computation. Acceptability is impaired due to unconsciousness or reluctance of participants. Group evaluation overcome such problematic inputs from untrustworthy participation.

## Conclusion

A scarcity of adequate methodology still exists in promoting citizen participation in Smart City. IoT enabled Smart-devices and personalized services opens horizon of innovation regarding qualitative experiences. Consequently, context awareness is deemed both opportunity and pitfall for overall acceptability of services. It leverages plethora of monitoring information for the systematic realization of groundbreaking urban services. However, the heterogeneity in monitoring data introduces composite relationship between contexts and participation. Ours is a step in harnessing the blue ocean, while meeting major pitfalls. In a nutshell, our non-parametric Bayesian scheme addresses complex interactions among contexts, whilst scaling up over time. Consequently, systematic prototype is implemented for the accurate prediction of engagement by assessing participants' opinions, usages or even feed-backs.

**Acknowledgment** The manuscript is submitted as part of IEEE, ACM Voluntary and Professional services.